\documentclass[5p,authoryear,preprint]{elsarticle}

\usepackage{amssymb,lineno,hyperref,color}
\modulolinenumbers[5]

\journal{Astronomy and Computing}

\begin{document}

\begin{frontmatter}

\title{Modelling grain-size distributions in C-type shocks using a discrete power-law model}
\author[1]{R. Sumpter}
\ead{rosie.sumpter@gmail.com}
\address[1]{School of Physics and Astronomy, University of Leeds, Woodhouse Lane, Leeds LS2 9JT, UK}

\author[1,2]{S. Van Loo\corref{cor1}}
\ead{Sven.VanLoo@UGent.be}
\cortext[cor1]{Corresponding author}
\address[2]{Department of Applied Physics, Ghent University, Sint-Pietersnieuwstraat 41, Technicum blok 4
9000 Gent, Belgium}




\begin{abstract}
In this paper we discuss the implementation of a  discrete,  piecewise power-law grain-size distribution
method into a numerical multifluid MHD code as described in \citet{Sumpter2020}. 
Such a description allows to capture the full size range of dust grains and their 
dynamical effects. The only assumptions are that grains within a single discrete bin 
have the same velocity and charge. We test the implementation by modelling 
plane-parallel C-type shocks and compare the results with shock models of 
multispecies grain models. We find that both the discrete and  multispecies grain 
models converge to the same shock profile. However, the convergence for the discrete 
models is faster than for the multispecies grain models. For the pure advection 
models a single discrete bin is sufficient, while the  multispecies grain models need 
a minimum of 8 grain species. When including grain sputtering the necessary number 
of discrete bins increases to 4, as the grain distribution cannot be described by a 
single power-law as in the advection models. The multispecies grain models still 
need more grain species to model the distribution, but the number does not increase 
compared to the pure advection models. Our results show that modelling the grain 
distribution function using a discrete distribution reduces the computational cost
needed to capture the grain physics significantly. 

\end{abstract}

\end{frontmatter}

\section{Introduction}
Although the typical grain-size distribution in the Interstellar Medium (ISM) follows the  \citet{MRN77} (MRN) distribution, 
local variations are expected as the grain-size distribution is subjected to different grain processes such as dust production
 \citep{Maercker18}, grain growth by coagulation and  mantle accretion \citep{jones85,liffman89,ossenkopf93,inoue03,ormel09,asano13,ysard16,jones17}, 
 and destruction processes like sputtering, shattering and vaporisation  \citep{tielens94, jones96, flower03, hirashita09, guillet07, guillet09, guillet11, anderl13}.  These grain processes do not only affect the overall grain distribution, but also have significant effects on the dynamics of the ISM.  In the outflows of Young Stellar Objects (YSOs), dust grains are important charge and current carriers and therefore determine the structure of C-type shocks \citep{VanLooetal2009}. Furthermore, grain-processing leads to observational signatures. In typical ISM conditions silicon is adsorbed onto dust grains. However, gas phase SiO is detected in the clumpy structure of YSO outflows
due to the shock-induced sputtering releasing silicon into the gas phase \citep[e.g.][]{Martin-Pintadoetal1992, Mikamietal1992}. Thus, it is essential to accurately follow the evolution of the dust grain distribution to model both the dynamics and emission signatures.

A number of different approaches have been used to model the dynamics and processing of dust grains in C-type shocks. In the simplest method, i.e. the multispecies grain model,  only a few dust species with specified radii, i.e. typically, one or two grain species representing small and/or large dust grains, 
are evolved \citep[e.g.][]{DRD83,VanLooetal2009,vanloo13}. This approach captures not only the dynamical importance of dust grains
to the shock structure, but also the sputtering process of the grains. However, the limited number of dust species fails to represent the 
grain-size distribution adequately and is furthermore restricted in use as it cannot model grain-grain processes such as shattering 
and vaporisation \citep[][]{Sumpter2020}. A more rigorous approach uses a discrete distribution function in which the grain-size 
distribution is modelled using different size bins spanning the full radial range \citep[e.g.][]{jones96,guillet07,guillet09,guillet11,anderl13}.
While both sputtering and grain-grain processing are followed to provide a more realistic shock and emission profile,  these models
only follow the total dust mass and approximate the dust number density. \citet{mckinnon18} improved this method by using a linear
piece-wise discretisation in a grain size bin to conserve both mass and number density and showed it to be second-order accurate in 
number of size bins. However, this still requires 50-100 bins to achieve an accuracy of the order of 1\%. 
A further improvement was suggested by \citet[][hereafter Paper I]{Sumpter2020}: they show that a power-law discretisation of the distribution function preserves the global distribution properties to a high degree even with small bin numbers.

In this paper we will implement this discrete power-law method in a multifluid magnetohydrodynamics (MHD) code (Sect.~\ref{sect:numerical model}). Then, in Sect.~\ref{sect:advection}, we will compare the profiles of oblique C-type shocks obtained with both the power-law discretisation and multispecies approaches, while, in Sect.~\ref{sect:sputtering}, we will present the results for the sputtering of grain cores with the aim of determining the number of size bins required to accurately model this destructive grain sputtering process. Finally, Sect.~\ref{sect:conclusions} contains a discussion of the results and conclusions.

\section{Numerical model}\label{sect:numerical model}

\subsection{Dynamics of discrete grain distributions}

To model a weakly-ionised dusty plasma, we use the multifluid MHD code MG with the chemical network described in \citet[][2013]{VanLooetal2009}. The numerical scheme underlying the code solves the continuity, momentum and 
energy equation for the neutral particles.  In the limit of small mass densities for the charged fluids, their inertia 
can be neglected and, also, the charged fluids are in thermal balance as their heat capacities are small. Therefore we 
can restrict ourselves to solving the continuity and reduced momentum equations for all charged fluids and the reduced energy 
equations for ions and electrons only (as the dust grains can be treated as a pressureless fluid).
Furthermore, the neutral and charged fluids are coupled by mass, momentum and energy transfer coefficients. 
These equations are solved 
using a second-order hydrodynamic Godunov solver for the neutral fluid equations.  The charged fluid densities are calculated 
using an explicit upwind approximation to the mass conservation
equation, while the velocities and temperatures are calculated iteratively from the reduced momentum and energy
equations. The magnetic field is calculated using the induction equation with the resistivities obtained from Ohm's law.
The magnetic field is advanced explicitly, even though this implies a restriction on the stable
time step at high numerical resolution \citep{Falle2003}.

However, we make three modifications. While \citet{vanloo13} modelled grain sputtering and thus grain mass loss, the mass loss was not taken into account when determining the grain dynamics. Our first modification is to relax this assumption. The second change concerns the dust grain's charge. While in previous models the grains are always assumed to be negatively
charged, this is also relaxed and, in specific conditions, grains can attain a positive charge. The final change relates to the calculation of the velocities of the charged particles from the reduced momentum equations and provides a more robust derivation of these. The latter two modifications are described in \ref{app:A} and \ref{app:B}.

When incorporating the grain-size distribution in the multifluid MHD scheme the equations governing the (pressureless) grain fluids 
need to be modified. As the description of the implementation is detailed in Paper I we only give a short summary here. 
We assume that the grain number density distribution in bin $i$, given by the size interval $[a_i, a_{i+1}]$, has a power-law shape, i.e.
 \begin{equation}
 \left.\frac{\partial{n(a, t)}}{\partial{a}}\right|_i = A_i a^{-\alpha_{i}},
 \end{equation}
 where  $A_i$ is the power-law coefficient and $\alpha_i$ the power-law index which both implicitly depend on time.  Then, by integrating the continuity equations and reduced momentum equation over the grain number density distribution in bin $i$, the governing equations for bin $i$ become
\begin{eqnarray}
&\frac{\partial n_{i}}{\partial t} + \nabla \cdot \left( n_{i} \bar{{\bf v}}_i\right) = S_{i,{\rm sputt}},\\
&\frac{\partial \rho_{i}}{\partial t} + \nabla \cdot \left(\rho_{i} \bar{{\bf v}}_i\right) =  S'_{i, {\rm sputt}},\\
&\langle Z\rangle_i e n_i \left({\bf E} + \bar{{\bf v}}_i \times {\bf B}\right) + n_i \rho_{n} K^*_{gn} ({\bf v}_n - \bar{{\bf v}}_i) = 0,\label{eq:redmom}
\end{eqnarray}
where $n_i$ is the total number density in bin $i$, $\rho_i$ the total mass density, and $\langle Z \rangle_i e$ the average grain charge. Furthermore, 
\begin{equation}\label{eq:K_gn}
 K^*_{gn} = \frac{8\pi}{3} \langle a^2\rangle_i \left(\frac{2k_BT_n}{\pi m_n}\right)^{1/2} 
 \left(1 + \frac{9\pi ({\bf v}_n - \bar{{\bf v}}_i)^2}{128k_BT_n}\right)^{1/2},
\end{equation}
is the mean specific collision coefficient between neutrals and grains in bin $i$ (and is determined by  integrating the collision coefficient $K_{gn}$). Here, the average value of a variable $f$ is given by
\begin{equation}
	\langle f \rangle_i = \frac{1}{n_i}\int\limits_{a_i}^{a_{i+1}}{da\ f \left.\frac{\partial n}{\partial a}\right|_i}, 
\end{equation}

The only assumption made in the derivation of these equations is that all the grains within a bin have the same velocity ($\bar{\mathbf{v}}_{i}$). While this is not necessarily
valid, it is expected to only have a minor effect. The grain velocity depends effectively on the grain radius through the Hall parameter, which is the ratio of the gas-grain collision frequency to the gyrofrequency, $\beta = ZeB/(m\rho_{n}K_{gn}) \propto a^{-1}$. For small Hall parameters, i.e. when $|\beta| < 0.1$, the grains move with the neutrals, while, for $|\beta| > 2$ they move with the electrons and ions. Thus, there is only a small range of $\beta$ values, or grain radii, for which grains have a velocity in between these limits. 

 $S_{i,{\rm sputt}}$ and $S'_{i,{\rm sputt}}$ represent the sputtering losses in bin $i$ and are calculated as described in Paper I. For this, we specifically
need to calculate the rate of change of the grain radius, i.e. $\dot{a}$, in each bin due to sputtering. The sputtering yields due to H$_2$, CO, O, Mg, H$_2$O  and SiO projectiles are calculated using the same expressions as in \citet{vanloo13}. It is important to realise that $\dot{a}$ is not explicitly dependent  on the grain size, only implicitly via the grain-neutral relative velocity. Given that the velocity is assumed to be the same for all grains in a bin, this relative velocity, and thus also $\dot{a}$ is also constant in each bin (which is implicitly assumed in Paper~I).

\subsection{Initial conditions}
We follow \citet{VanLooetal2009} and \citet{vanloo13} for the initial set-up of the shock models. This assumes that the initial flow profile corresponds to 
that of a discontinuity similar to a J-type shock.  The upstream magnetic field is $B_0 = 1\mu G  (n_{H}/1~{\rm cm^{-3}})^{1/2}$, where $n_{H}$ is the upstream value of the number density of hydrogen nuclei. A shock propagates with a speed $v_s$ at an angle $\theta = 45^\circ$ with respect to the upstream magnetic field. We will consider two different shock set-ups corresponding to models with or without grain sputtering.
For the advection only models  (no sputtering), the  pre-shock density is either $n_{H} = 10^{4}~{\rm cm^{-3}}$ or $n_{H} = 10^{6}~{\rm cm^{-3}}$ with 
a shock speed $v_s = 25~{\rm km~s^{-1}}$, while, for the sputtering models, the shock propagates with a higher speed of $40~{\rm km~s^{-1}}$. We adopt this larger shock speed for sputtering as $25~{\rm km s^{-1}}$ is the threshold speed for core sputtering and would 
have little effect on the dust distribution.  Furthermore we adopt the same fractional abundances for the upstream chemical species, that is the upstream gas phase
fractional abundances of O, Mg, H$_2$O and SiO are  $4.25\times 10^{-4}$, $10^{-7}$, 0 and 0, respectively. The fractional abundance
of CO we keep constant at $5\times 10^{-5}$ throughout the numerical domain.

The dust grains are modelled using an initial power-law size distribution of grain cores with an index of -3.5 \citep[][referred to as the MRN distribution function]{MRN77}  and set the lower and 
upper radial limit of the distribution to $a_{\min} = 5\times10^{-7}$~cm and 
$a_{\max} = 3\times10^{-5}$~cm, respectively. We do not include grain mantles in this paper as we
consider a shock speed above $\approx 20 {\rm km~s^{-1}}$ for which grain mantles are completely, but 
also rapidly,  eroded \citep{guillet11,vanloo13}.  
In the upstream region, one percent of the mass is contained in grains.  The full radial range of the initial distribution is divided up logarithmically 
with a spacing determined by 
\begin{equation}
\Delta a = \frac{\log(a_{\max}/a_{\min})}{N},
\end{equation}
where $N$ is the number of bins. Then the  edges of bin $i$ are effectively $a_i = a_{\min} e^{i\Delta a}$ and $a_{i+1} = a_{\min} e^{(i+1)\Delta a}$.
The number density $n_i$ in each bin $i$ can then be calculated as
\begin{equation}
	n_i = \int\limits_{a_i}^{a_{i+1}}{da\ \left.\frac{\partial n}{\partial a}\right|_i}, 
\end{equation}
and the mass density $\rho_i$ as
\begin{equation}
\rho_i = \frac{4\pi}{3} \rho_g n_i \langle a^3\rangle_i,
\end{equation}
where $\rho_g$ is the density of the grain material. For models with the multispecies grains, an appropriate radius $\bar{a}_i$ needs to chosen for each bin. 
As the average mass per grain is given by $\rho_i/n_i$, it is clear that $\bar{a}_i = \left(\langle a^3\rangle_i \right)^{1/3}$.

\section{Advection of the grain distribution}\label{sect:advection}
While Paper~I studied a numerical approach for accurately evolving a dust grain-size distribution undergoing grain processing, it only suggested 
an implementation of the technique in a (magneto)-hydrodynamical code. Therefore, it is yet unclear whether a discrete distribution function can
adequately be used for modelling the dynamics. Here we compare the shock structures using both single-species and discrete distributions with 
the number of species (or bins)  ranging from 1 to 16.

\begin{figure}
\includegraphics[angle=-90,width=\columnwidth]{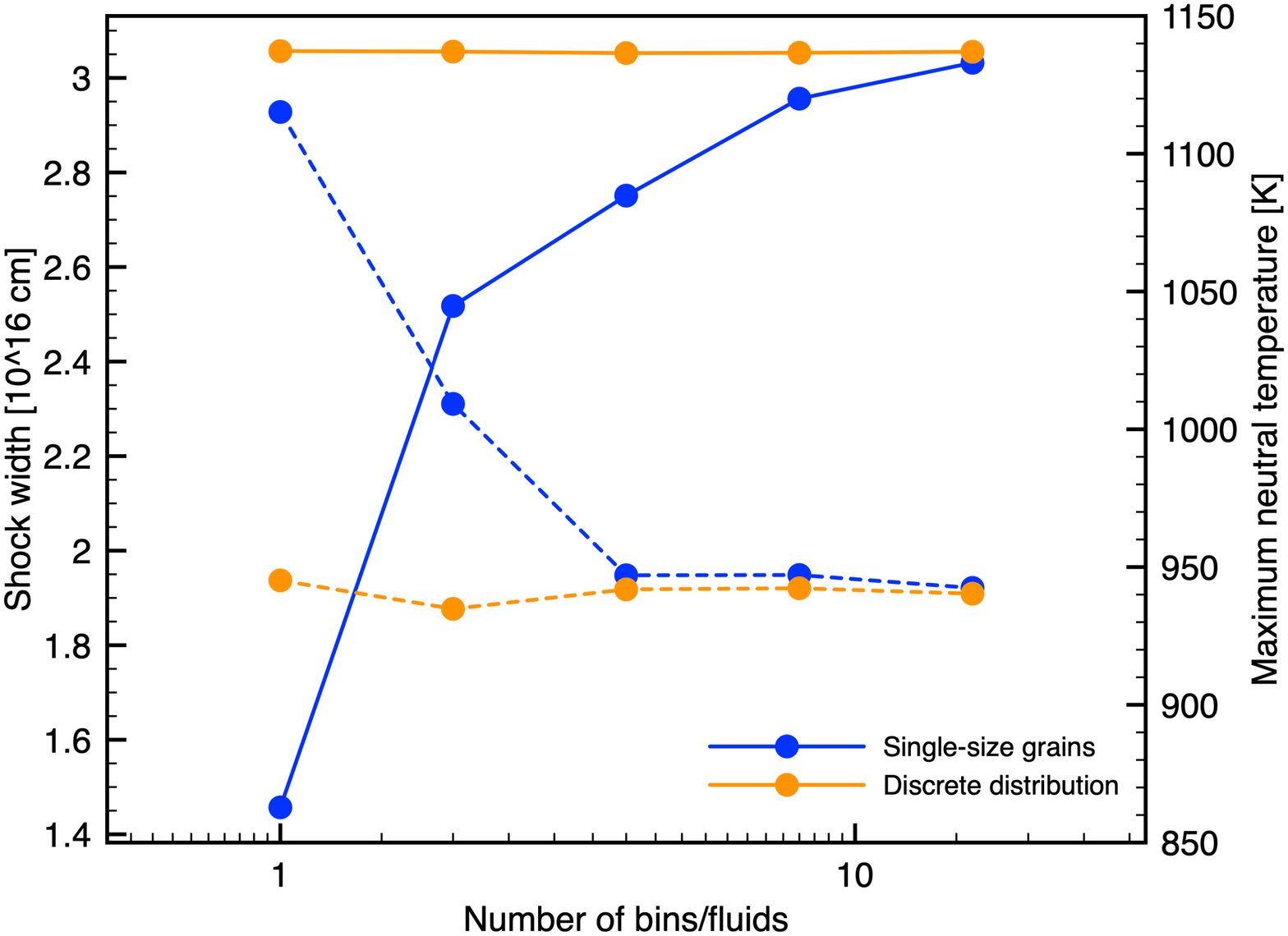}
\caption{Shock width (solid line) and maximum neutral shock temperature (dashed line) as function of 
number of multispecies grains (blue) or discrete bins (amber) for $v_s = 25~{\rm km s^{-1}}$ and 
$n_H = 10^4~{\rm cm^{-3}}$.} 
\label{fig:convergence_advection_nh4}
\end{figure}

\begin{figure}
\includegraphics[angle=-90,width=\columnwidth]{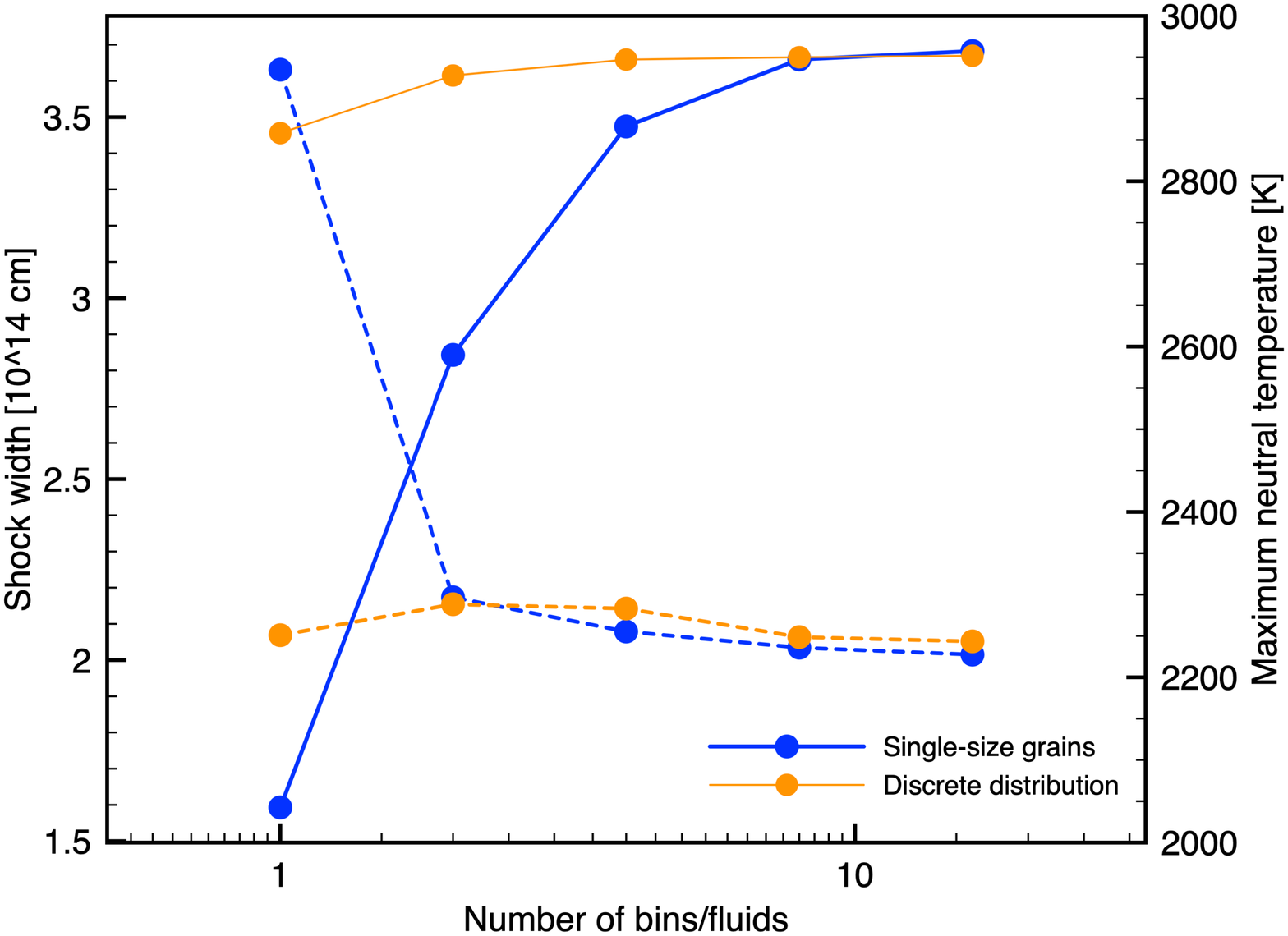}
\caption{Shock width (solid line) and maximum neutral shock temperature (dashed line) as function of
number of multispecies grains (blue) or discrete bins (amber) for $v_s = 25~{\rm km s^{-1}}$ and
$n_H = 10^6~{\rm cm^{-3}}$.} 
\label{fig:convergence_advection_nh6}
\end{figure}

Figures~\ref{fig:convergence_advection_nh4} and \ref{fig:convergence_advection_nh6} show characteristic measures for the shock structure:  
the shock width measured as the part of the shock with a neutral temperatures above 100~K \citep[as in][]{anderl13} and the maximum neutral temperature
as hotter temperatures are achieved in thinner shocks with a narrower radiating layer. We find that the shock is thinnest (2-2.5 times) and hottest (25\%) 
for the model with one grain species. As more multispecies grains are included, the shock width and maximum temperature seem
to converge. This convergence is more apparent if we compare the results with the ones from the discrete distribution function.  Both models tend to the same values.
However, contrary to the multispecies grain models,  the shocks widths and maximum temperatures for the discrete distribution models have quasi-constant values even for a low number of bins. For the $n_H=10^4{\rm cm^{-3}}$ runs the relative variation in the shock width is less than 1\%, while it is slightly higher for the  $n_H=10^6{\rm cm^{-3}}$ model at 6\%, compared to  52\% and 56\% for the multispecies grain models respectively. 

To understand this behaviour, we need to understand the physical process governing the shock 
structure. In particular, the shock width is set by balancing the Lorentz force with the 
drag forces. From summing the reduced momentum equations (Eq.~\ref{eq:redmom})  for the charged particles together, 
we find
\begin{equation}\label{forcebalance}
{\bf j} \times {\bf B} + \rho_n \sum_i{\rho_i K_{in} ({\bf v}_n - {\bf v}_i)} = 0.
\end{equation}
Although dust grains are not  necessarily the dominant charge carriers (they are 
for $n_H = 10^6~{\rm cm^{-3}}$, but not for $n_H = 10^4~{\rm cm^{-3}}$), they do contribute 
significantly to the drag force. In regions with low ionisation fraction 
$\chi$, i.e. $\chi < 3\times 10^{-9} (v_s/{\rm km~s^{-1}})$, as in these models, 
the drag force of the neutral-grain collisions dominates the ion-neutral drag force 
\citep{VanLooetal2009}. As the neutral-grain drag force is proportional to $\rho_g K_{gn}$, 
where $\rho_g$ is the grain mass density, the smaller grains in a MRN size 
distribution contribute the most to the drag. 

\begin{figure*}
\centering
\includegraphics[angle=-90,width=\textwidth]{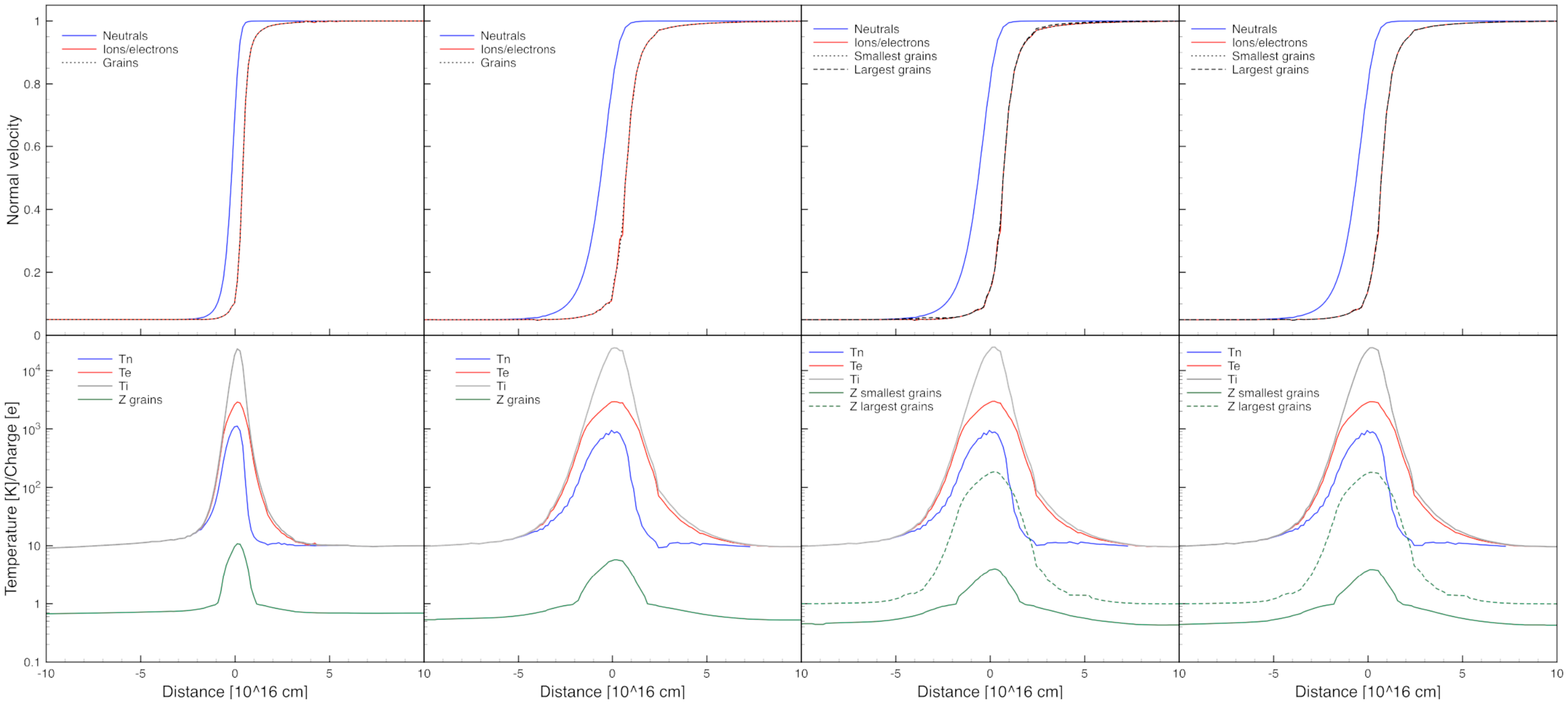}
\caption{Shock structure for different models of a shock propagating at $25~{\rm km~s^{-1}}$ through $n_H = 10^4~{\rm cm^{-3}}$. From left to right: 1 grain species, 1 discrete bin, 16 grain species and 16 discrete bins. Top panel shows the normal velocity in the shock frame of the neutrals, ions/electrons and grains normalised to the shock velocity, while bottom panel show neutral, electron and ion temperatures as well as the absolute value of the grain charge.} 
\label{fig:adv_4_all}
\end{figure*}

\begin{figure*}
\centering
\includegraphics[angle=-90,width=\textwidth]{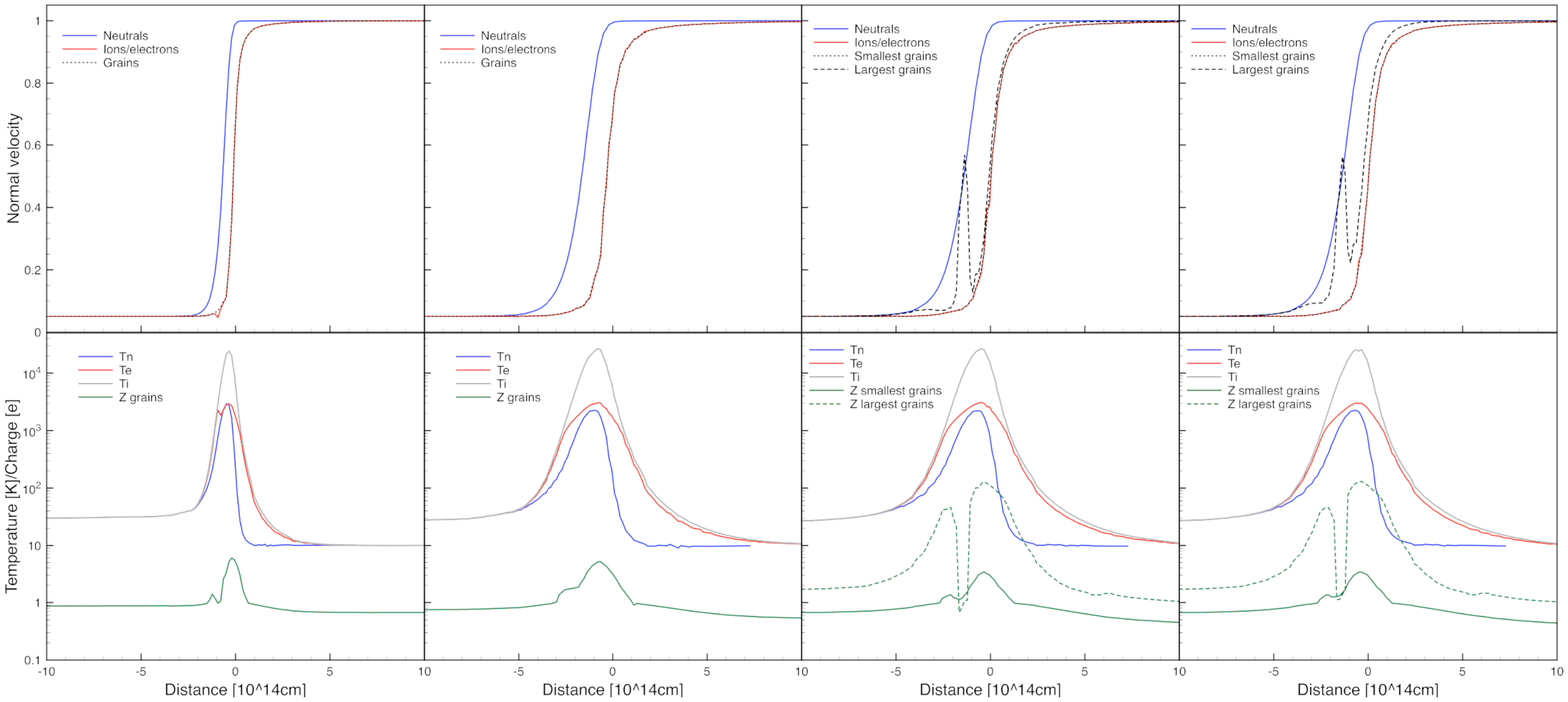}
\caption{Shock structure for different models of a shock propagating at $25~{\rm km~s^{-1}}$ through $n_H = 10^6~{\rm cm^{-3}}$.
From left to right: 1 grain species, 1 discrete bin, 16 grain species and 16 discrete bins. Top panel shows the normal velocity in the shock frame of the neutrals, ions/electrons and grains normalised to the shock velocity, while bottom panel show neutral, electron and ion temperatures as well as the absolute value of the grain charge. }
\label{fig:adv_6_all}
\end{figure*}

In the discrete distribution models, the collision coefficient is averaged over all of the grain 
radii within a bin (see Eq.~\ref{eq:K_gn}) so that 
the evaluation of the drag force only depends on the number of bins needed to adequately 
describe the distribution function. 
For $n_H = 10^4~{\rm cm^{-3}}$ all grains move with the electrons and ions as seen in 
Fig.~\ref{fig:adv_4_all}. This means that the dust distribution keeps the same initial power-law 
distribution as the grains move from upstream to downstream and can, thus, be described by 
single discrete bin. However, the $n_H = 10^6~{\rm cm^{-3}}$ models show a more complicated 
situation.  The higher neutral density implies an increased collision frequency and the 
larger grains become decoupled from the electrons and ions (see Fig.~\ref{fig:adv_6_all}) 
and even move along with the neutrals for a short distance.  This is because, for 
$n_H = 10^6~{\rm cm^{-3}}$, the grains become the dominant charge carriers depleting the
available free electrons that attach to the grains causing the largest grains to become 
quasi-neutral. Consequently, their Hall parameter is smaller than unity and the grains 
start moving with the neutrals so that these grains do not contribute to the drag force. 
The distribution function then needs to be modelled using more than 1 discrete bin. However,
as the contribution of the large grains to the dust drag is limited, the overestimation of 
the drag force is small as can be seen in Fig.~\ref{fig:convergence_advection_nh6}.

For the single-size models the collision coefficient is not evaluated across a range of 
grain radii, but just at $\bar{a}_i$. Remember that this value is determined from the 
initial distribution function (see Sect.~\ref{sect:numerical model}). When considering
only one grain species or discrete bin, it is possible to compare the drag force exerted by 
the dust, i.e. the ratio of the single size to discrete bin model is given by 
$\rho_g K_gn/n_g K^*_{gn} = \langle a^3 \rangle^{2/3}/\langle a^2 \rangle$ where the average is determined across the 
full grain size range $[a_{\rm min}, a_{\rm max}]$. For an initial MRN distribution 
this ratio is then $\approx 2.4$ which also is the relative difference observed in the shock widths 
between the single size and discrete bin models in Figs.~\ref{fig:convergence_advection_nh4} and 
\ref{fig:convergence_advection_nh6}. Including more grain species will reduce this overestimation 
in the drag force until convergence. We find that at least 8 to 16 grains species are required for the results to converge. 

Both the discrete models and the multispecies grain models converge to the same values giving confidence that the discrete distribution models reproduce the correct shock structures. The importance of using a discrete distribution of grain sizes is demonstrated, with multispecies fluids only accurately achieving the same shock profile as the distribution for upwards of about 8 fluids. While discrete models with a single bin produce good results, it is useful to use a minimum 
of two bins to model the different dynamics of small and large grains. This becomes especially important when including grain processing physics such as sputtering in the next section.

\section{Sputtering of the grain distribution}\label{sect:sputtering}
Here we explore the effect of sputtering as the initial MRN distribution is advected through the shock front. In the previous section we found that, for advection-only models, 
only two discrete bins are necessary to correctly simulate the shock dynamics. However, grain 
sputtering depends on the relative velocity between the grains and the 
impacting particles which can either be ions or neutrals. Therefore it is not only important to describe the grain size distribution but also the velocity distribution. As the grain velocity in a size bin is assumed to be constant, the velocity distribution of the grains 
is better described by more size bins as will be the sputtering of those grains. 

\begin{figure}
\includegraphics[angle=-90,width=\columnwidth]{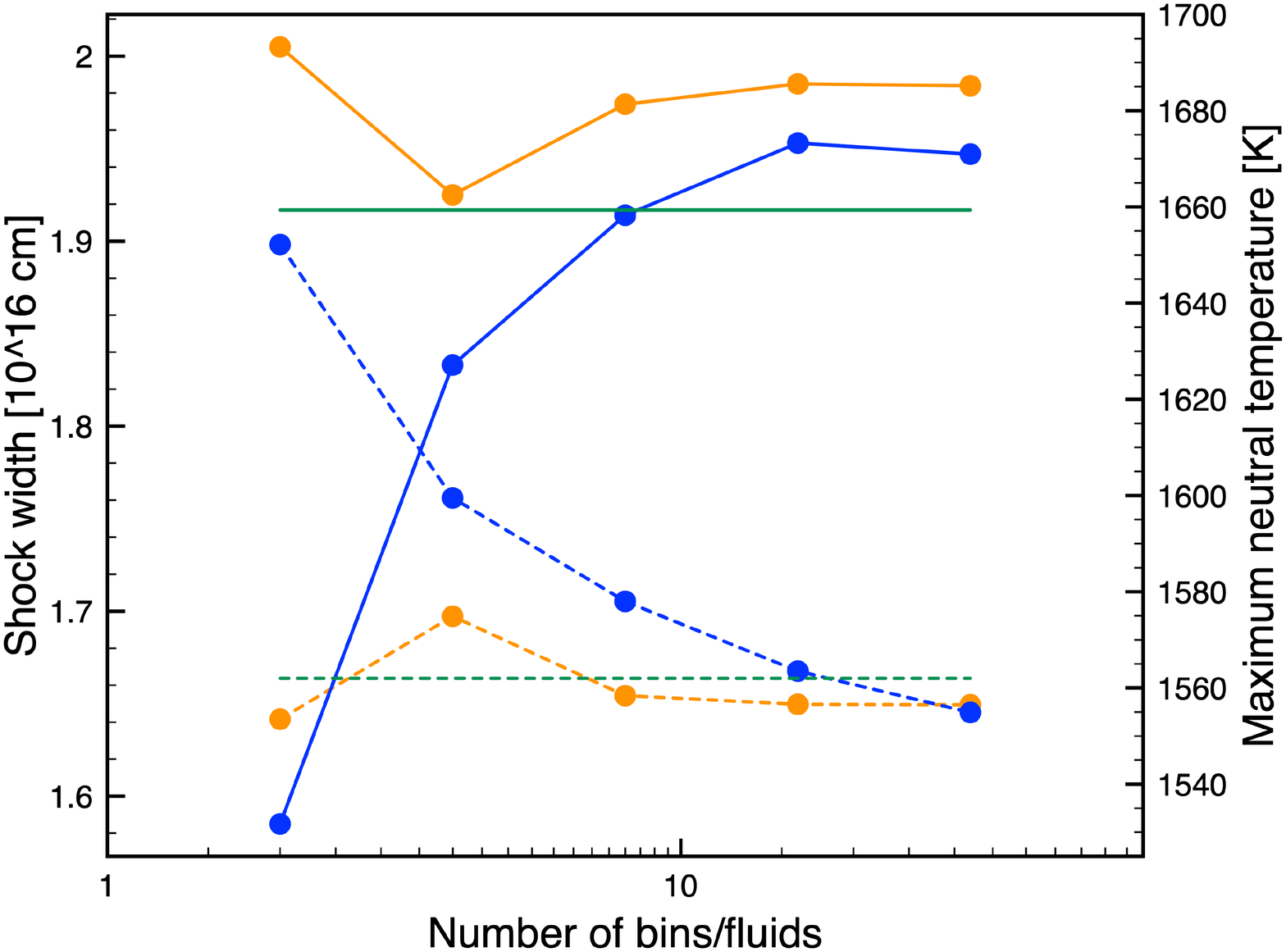}
\caption{Shock width (solid line) and maximum neutral shock temperature (dashed line) as function of
number of multispecies grains (blue) or discrete bins (amber)
for $v_s = 40~{\rm km s^{-1}}$ and 
$n_H = 10^4~{\rm cm^{-3}}$ and including sputtering. The green lines show the converged shock 
properties for the same model without sputtering.} 
\label{fig:convergence_sputter_nh4}
\end{figure}

\begin{figure}
\includegraphics[angle=-90,width=\columnwidth]{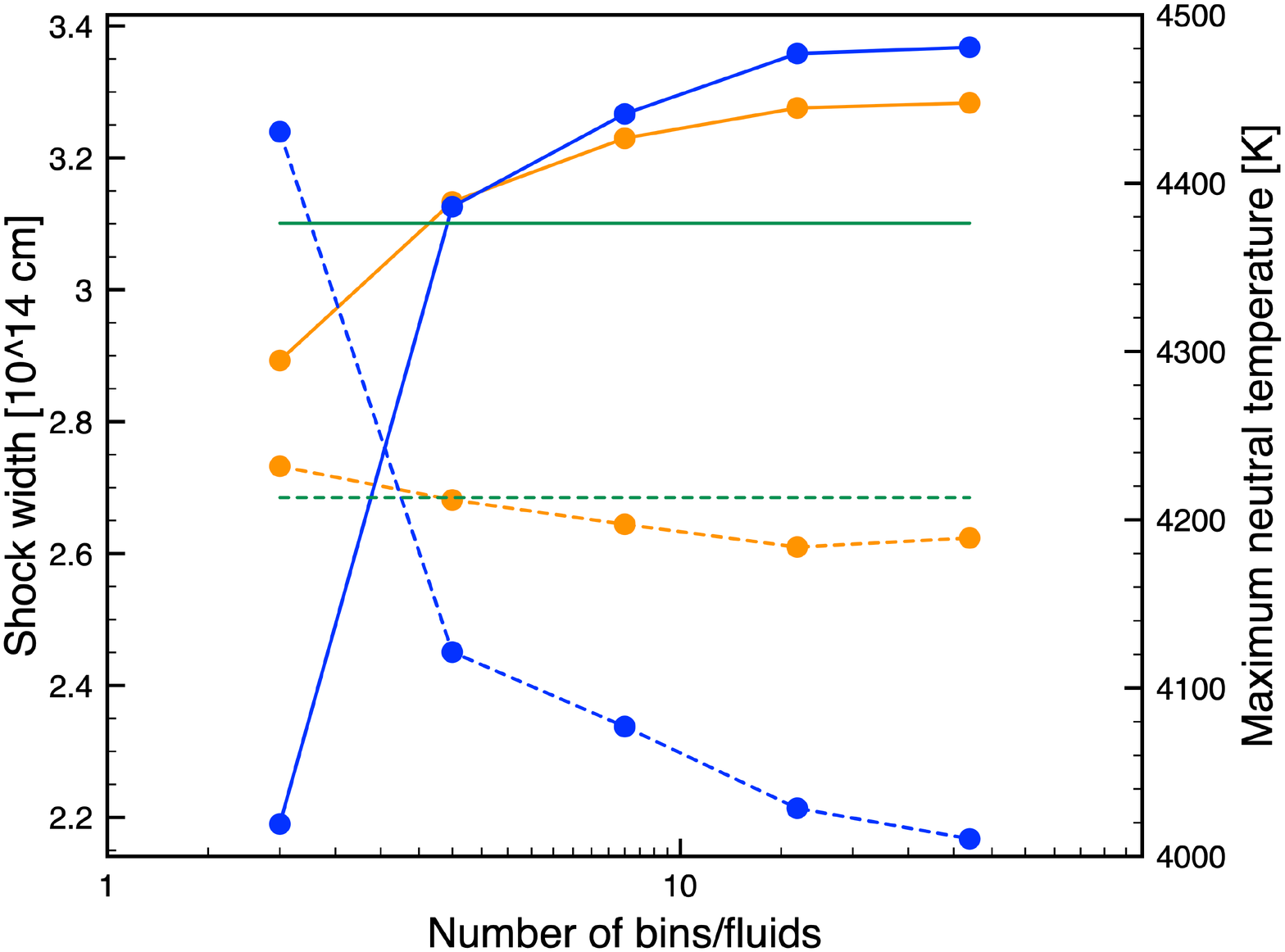}
\caption{Shock width (solid line) and maximum neutral shock temperature (dashed line) as function of
number of multispecies grains (blue) or discrete bins (amber) for $v_s = 40~{\rm km s^{-1}}$ and 
$n_H = 10^6~{\rm cm^{-3}}$. The green lines show the converged shock
properties for the same model without sputtering.} 
\label{fig:convergence_sputter_nh6}
\end{figure}

\begin{figure}
\includegraphics[angle=-90,width=\columnwidth]{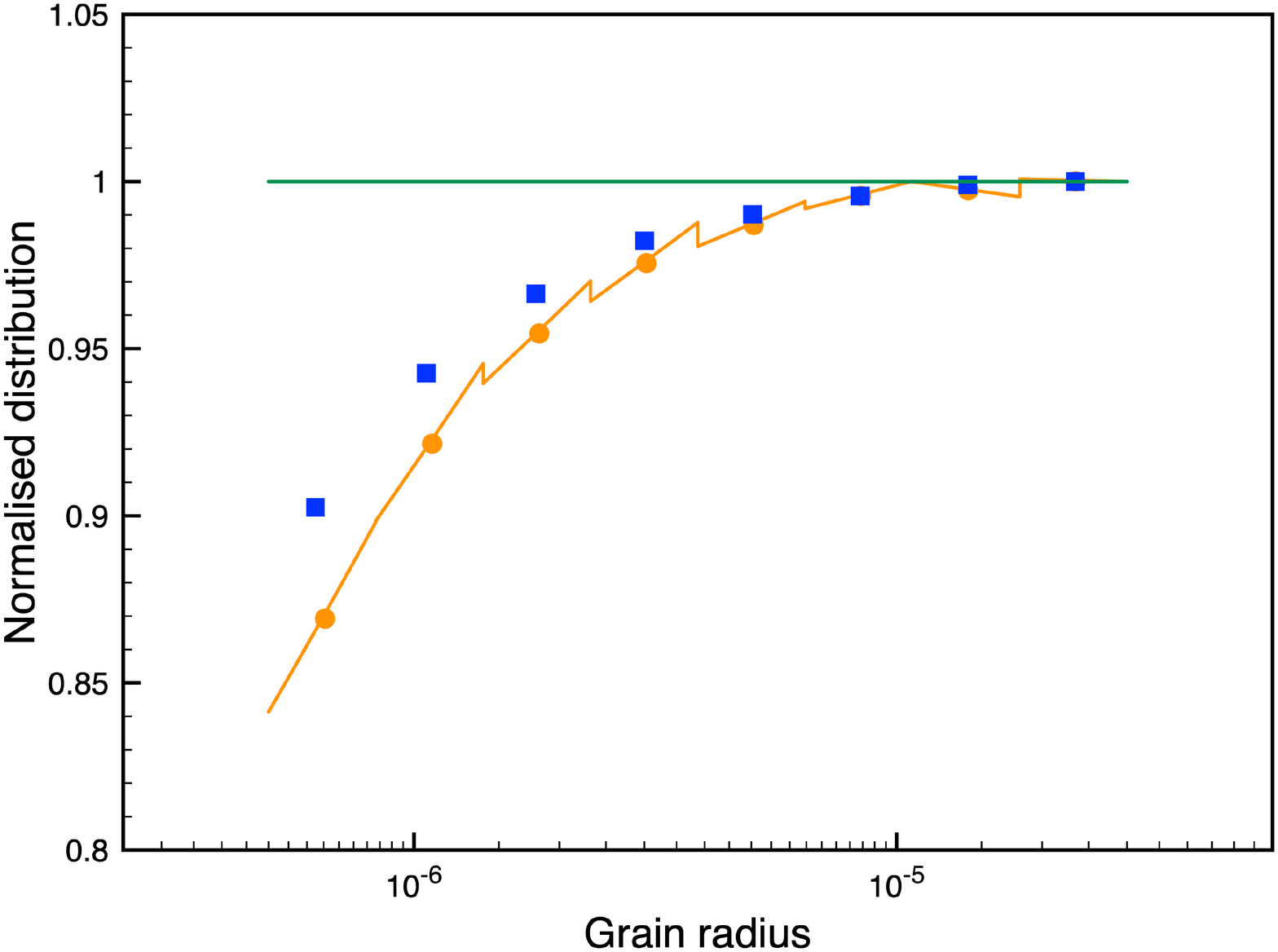}
\caption{Grain size distribution normalised to the MRN distribution for both the discrete (amber) and 
single-size (blue) models for $v_s = 40~{\rm km s^{-1}}$ and $n_H = 10^4~{\rm cm^{-3}}$ with $N=8$. 
The symbols show the total number density of each bin or fluid (i.e. $a^{2.5} n_i$). The values for 
the discrete bins are positioned at the average grain radius of the bin. The solid lines show the discrete distribution function (amber) and the MRN distribution (green) (i.e. $a^{3.5}\frac{dn}{da}$).}
\label{fig:distribution_nh4}
\end{figure}

Figures~\ref{fig:convergence_sputter_nh4} and \ref{fig:convergence_sputter_nh6} show the shock properties for a $40~{\rm km s^{-1}}$ shock propagating in a 
$n_H = 10^4~{\rm cm^{-3}}$ and $10^6~{\rm cm^{-3}}$ medium as function of the number of discrete bins or multispecies while including grain sputtering.
Similarly to the advection models (Figs.~\ref{fig:convergence_advection_nh4} and \ref{fig:convergence_advection_nh6}) the shock width and the maximum neutral 
temperature converge as more discrete bins or multispecies are used to model the grain distribution. This is more so for the multispecies models than for 
the discrete bin models. For $n_H = 10^4~{\rm cm^{-3}}$, the multispecies models have a relative variation in the shock width from 19\% for $N=2$ and dropping below 
5\% for $N=8$, while the discrete bin models are all within 2\%. For $n_H = 10^6~{\rm cm^{-3}}$,  the relative differences are larger as in the pure-advection models: 
34\%  and 12\% for the multispecies and discrete distribution models respectively when $N=2$, but  fall below 5\% for $N=4$ for the discrete models and $N=8$ for the 
multispecies models. Similarly the maximum neutral temperature exhibits the same convergent behaviour although the variation is smaller, i.e. all relative differences are 
smaller than 5\% except for the multispecies grains for $N=2$ when the variation is 6\% and 10\% for $n_H = 10^4~{\rm cm^{-3}}$ and $10^6~{\rm cm^{-3}}$ respectively.
Note that the maximum neutral temperature for the $n_H = 10^6~{\rm cm^{-3}}$ models is above 4000~K which is when $H_2$ dissociation  becomes important 
\citep{DRD83}. Then molecular cooling is significantly reduced and it is unlikely that a C-type shock can exist in these conditions. However, as we do not include 
$H_2$ dissociation, the transition from C-type to J-type shock does not naturally occur in our models. When including sputtering, more discrete bins are needed 
before convergence of the shock properties. Furthermore, the multispecies models still need to include more multispecies grains than bins needed in the discrete distribution 
models, i.e. $N=8$ compared to $N = 4$. 

\begin{figure}
\includegraphics[angle=-90,width=\columnwidth]{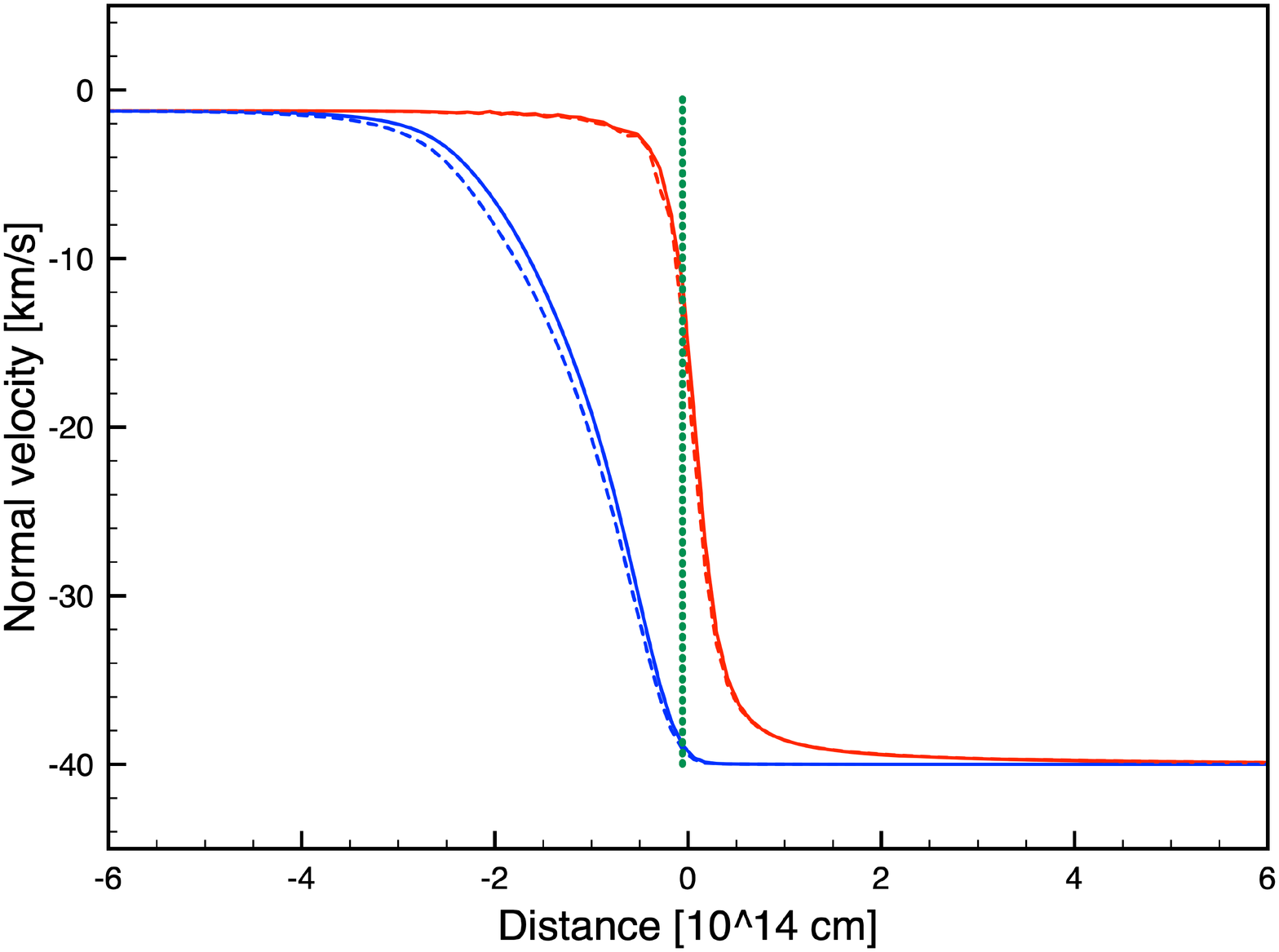}
\caption{Normal velocity in shock frame for a $v_s = 40~{\rm km s^{-1}}$ shock propagating in $n_H = 10^6~{\rm cm^{-3}}$. The solid line shows the ion velocity (red)
and neutral velocity (blue) for the model without sputtering, while the dashed line shows the model with sputtering. The green dotted line shows the position where 
the relative velocity difference between ions and neutrals is first $25~{\rm km s^{-1}}$.} 
\label{fig:diff_sp_nosp}
\end{figure}

When including sputtering the shocks widths are larger than for the advection-only models. This can be readily explained by the 
neutral-grain drag force: as the grains experience sputtering the grain distribution alters. At the small grain radii the distribution drops below the typical MRN distribution
function (see e.g. Fig.~\ref{fig:distribution_nh4}). As the smallest grains contribute most to the neutral-grain drag, the total drag force decreases. 
To maintain the force balance in Eq.~\ref{forcebalance} the shock width needs to increase (as ${\bf j} \times {\bf B} = (\nabla \times {\bf B}) \times {\bf B} \propto B^2/L$ where $L$ is the 
shock width). This can be also
inferred from Fig.~\ref{fig:diff_sp_nosp} which shows the normal velocity for the ion and neutral species for both advection and sputtering models.  While the ion velocity 
does not change between the models, the neutral velocity shows a slower deceleration in the shock frame. The deviation starts when the relative velocity between the 
ions and neutrals gets above $25~{\rm km s^{-1}}$. This corresponds to the threshold for grain core sputtering (most grains move with the ions and electrons). As 
the distribution changes due to sputtering and the grain-neutral drag reduces, the neutrals are not decelerated as quickly as in the advection only model.

Furthermore the neutral-grain drag force also explains the relative difference of 1-2.5\% in the shock width 
between the discrete distribution and  multispecies grain models contrary to the convergence seen in the advection only models. 
We see in Fig.~\ref{fig:distribution_nh4} that the multispecies distribution function has the same overall shape as the discrete 
power-law distribution function, but lies a bit higher. This is because, in our models, grains are assumed to be completely 
destroyed if their radius drops below $a_{\rm min} = 5\times 10^{-7}$. As the discrete distribution describes grains with 
a power-law size-distribution in each bin, dust grains with radii close to $a_{\rm min}$ will lose enough mass due to 
sputtering to reduce their radius below $a_{\rm min}$. In the  multispecies grain model with $N=8$ grain species, the 
smallest grains, with radius $\bar{a}_0$, do not experience enough sputtering for their radius to be reduced below 
$a_{\rm min}$. Hence, the number of grains remains constant in the multispecies models and decreases in the discrete model. 
Note that, if $N$ becomes large enough in the multispecies models, the smallest grains will be also removed from the 
distribution. This in return changes the amount of drag experienced between neutral and charged particles and changes 
the shock width. 

While we already discussed the fact that the dust grains are sputtered and that the grain-size distribution alters, we 
have not discussed how the sputtering process affects different grain sizes. This differs between the low and high 
density models. For the low density models, the grains contribute to the drag force, but do not carry much of the 
charge. As a result, all dust grains have a large Hall parameter and move with the other charged particles, i.e. the 
electrons and ions. As the relative velocity difference with the neutrals is then the same for all grains, the rate of 
change of the grain radius $da/dt$ is independent of grain radius. However, although all grains experience the same 
rate of change $\Delta a$, its effect is more readily seen at small grain radii as it is a larger fraction of their initial size. 
Figure~\ref{fig:distribution_nh4} shows this effect by normalising the distribution with the initial MRN distribution. 
While the large grains still follow the initial MRN distribution, the distribution at small grain radii has a smaller 
power-law index.  As seen before,  both the multispecies models and discrete bin
models produce similar results although the  multispecies models lie slightly higher.
Sputtering actually removes about 13\% of the total number of grains in the discrete model, while 
the multispecies models do not lose any. Thus the difference is mainly due to the normalisation with different total 
grain numbers.
However, we note that the mass lost in both the single-size and discrete distribution models
is similar at about 4\% as the small grains that are lost do not contribute much to the total grain mass.
For the higher density models,  the grains do carry a significant amount of the total charge. In fact, they dominate the free electrons in a large region of the shock front. 
Due to the lack of free electrons the large grains are close to zero or even positive charge and, thus, have a small Hall parameter. The large grains then move
predominantly with the neutrals. As the dominant sputtering projectile is $H_2$, the sputtering yield from the larger grains is non-existent. Only the small grains (up to 
$6\times 10^{-6}$cm) contribute to the sputtering and release of grain material such as silicon.  However, the resulting downstream grain distribution function has a similar 
shape as the one for $n_H = 10^4~{\rm cm^{-3}}$ (Fig.~\ref{fig:distribution_nh4}) with the large grains following the MRN distribution and a turn-over to lower power-law 
indices at small grain radii. So, although sputtering acts differently at the two densities, this cannot be inferred from the downstream distribution function.

\begin{figure}
\includegraphics[angle=-90,width=\columnwidth]{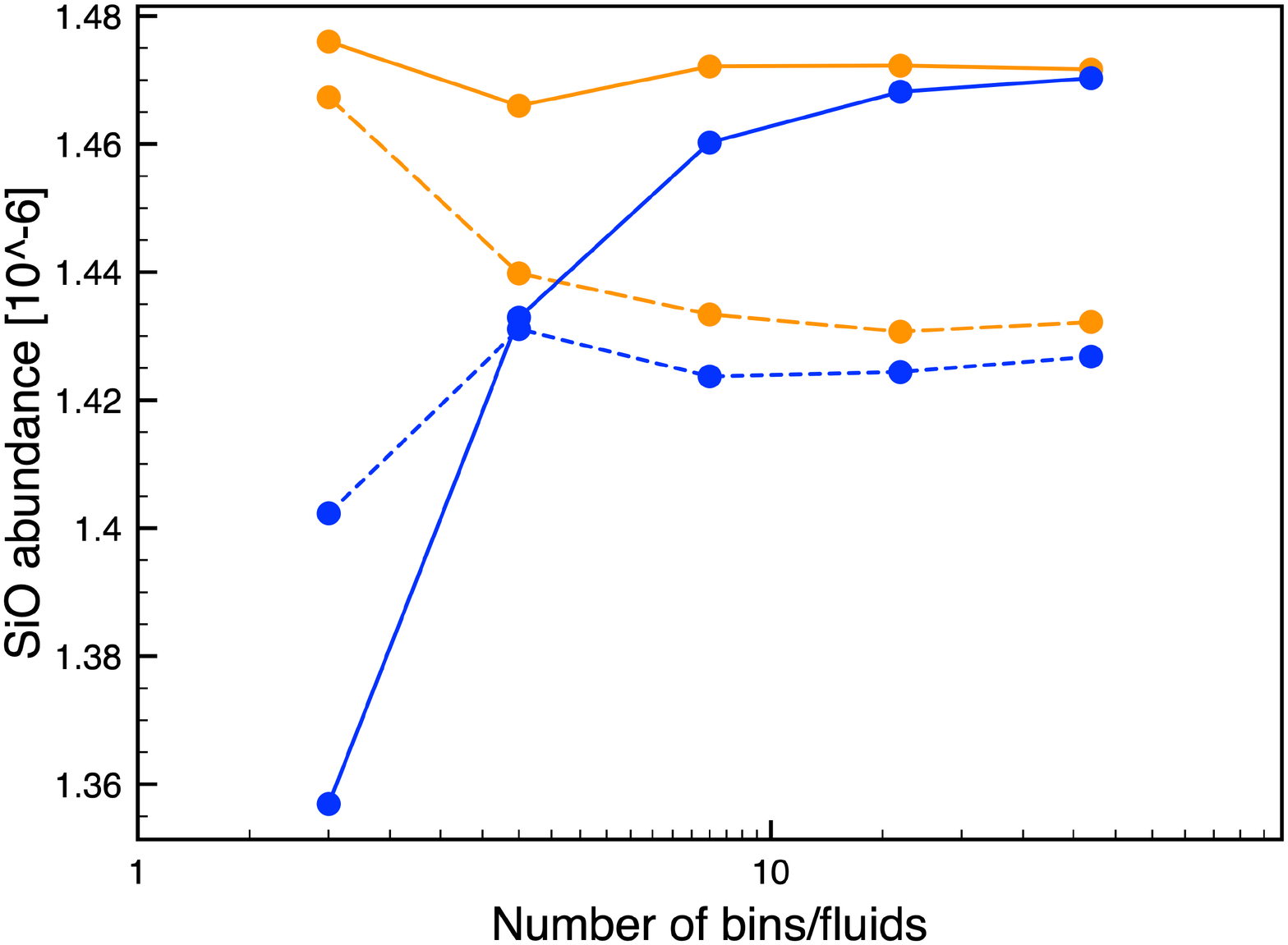}
\caption{SiO abundance for multispecies (blue) and discrete distribution (amber) models with $v_s = 40~{\rm km s^{-1}}$ (solid) and $n_H = 10^6~{\rm cm^{-3}}$ (dashed), when including sputtering.} 
\label{fig:convergence_SiO_nh6}
\end{figure}

As the dust grains loose mass due to sputtering, silicon returns to the gas phase as SiO. 
Figure~\ref{fig:convergence_SiO_nh6} shows the post-shock abundance of SiO for both background densities.  Again we notice a convergence of the values as the number of bins or species increases. However, the relative difference is quite small: the only relative difference above 5\% is when using a multispecies model  with $N = 2$ and even that is only 8\%. All other models have relative differences below 2\%. Note that the shape of the SiO abundance as function
of number of bins or fluids for $n_H = 10^4~{\rm cm^{-3}}$ follows that of the shock width, but does not for the $n_H = 10^6~{\rm cm^{_3}}$ models. This is because, as discussed
above, in the lower density model, all grains move together with the other charged particles and the sputtering region within the shock front is a fraction of the shock width. On the other
hand, for the higher density, the larger grains decouple from the magnetic field and move with the neutrals. With lower number of bins or fluids, it is not possible to correctly
capture the velocity distribution across the entire grain size distribution and the sputtering yield is either over- or underestimated and is no longer a constant fraction of the 
shock width as in the lower density models.

\section{Discussion and Conclusions}\label{sect:conclusions}
In this paper we  implemented a discrete grain-size distribution method into a numerical multifluid MHD code as described in Paper~I. Such a description allows to capture the full size range of dust grains and their dynamical effects. The only assumptions made are that grains within a single discrete bin have the same velocity and charge. We tested the implementation by modelling plane-parallel C-type shocks and compared the results with shock models of multispecies grain models. 

For models without grain processing the advection of the dust distribution function is to very good accuracy  modelled by a single discrete size bin even for models where the large and small grains have a different velocity.   The error in the characteristic measures is 1\% for $n_H = 10^4~{\rm cm^{-3}}$ and only 6\% for $n_H = 10^6~{\rm cm^{-3}}$. This is not surprising in the 
low density case as all grains have large Hall parameters and move with the same velocity as the ions and electrons throughout the shock front. However, for the high density case, the large grains move not with the electrons and ions. Then a single discrete
bin still produces very accurate shock profiles because the shock structure of C-type shocks is dominated by the dust-grain drag exerted by the small grains. Furthermore, the grain-size distribution does not vary much from the initial distribution as it is advected through the shock front. Contrary to the discrete bin models, shock profiles produced with multispecies grains cannot reproduce the shock structure with a single grain species. They actually produce much hotter and thinner shock structures. The models require the use of minimal 8 grain species for the shock profile to converge to the same result as the discrete bin models.

Following the pure-advection models we considered the effect of grain processing. In this paper we only take into account 
grain sputtering as the discrete bin models are directly comparable with multispecies grain models. We use the method 
described in Paper~I for number-conserving processes. Similarly to the pure-advection models, the discrete bin models converge
quicker than the multispecies grain models. As the grain sputtering changes the size distribution, especially at the smaller
size range, the distribution no longer can be described with a single bin, but has to be increased to about a minimum of 4 bins.
The main reason for this is that sputtering changes the grain size distribution throughout the shock as material is removed from the grains. However, this is still less than for the multispecies models which need a minimum of 8 species. 
The converged results of the  multispecies and discrete bin models are not identical but similar to a relative error of a few 
percent. This is because the sputtering process is treated slightly different in the two models, i.e. in the discrete bin models 
grains with radii below $a_{\rm min}$ are removed from distribution. 

We find that sputtering increases the shock widths of the C-type shocks as this grain process removes a fraction of the 
small grains and, thus, reduces the dust-grain drag balancing the Lorentz force. Furthermore, as material is removed from
the grains, SiO is released into the gas phase. While sputtering has a large effect on the shock properties, especially
for the multispecies grain models, the effect on the SiO abundance is much less. The relative error when modelling the 
distribution with $N=2$ discrete bins or multispecies grains is less than 10\%. Furthermore the fraction of Si removed from the grains is in agreement with the results of \citet{vanloo13} whose models did not include the mass loss of the grains in 
determining the shock structure. Because only a small part of the shock front contributes to sputtering and thus to releasing SiO,
the dynamical effects do not have a significant impact on the sputtering result. 

Our results show that the implementation of the dust distribution function within a multifluid MHD code modelling C-type shocks is 
successful as the piecewise power-law and multispecies grain models  converge. The only assumption that remains is that the velocity of the grains in each bin is equal. Therefore  this method can be readily applied to other problems in which single or multispecies dust models have been used such as in e.g. dust transport and evolution in galaxies and 
protoplanetary discs. Most studies in these research areas either follow the dust distribution evolution 
\citep{asano13,Aoyamaetal2018,Granatoetal2021} or the transport of a single dust species \citep{Kanagawaetal2018,HuBai2021}. 
However it is obvious that the dust distribution affects the dust dynamics and vice versa. Recent studies of planets 
in protoplanetary discs \citep[e.g.][]{Drazkowskaetal2019,Karlinetal2022}
reveal the interplay of multiple grain species on each others dynamics. Also, simulations of the streaming 
instability which plays a role in the early stages of planet formation show that, when  
multiple dust species are considered, the resulting instability growth rate depends strongly on the number of 
dust species considered \citep{Krappetal2019}. \citet{ZhuYang2021} show that, in the regime of slow growth rate, the number
of species required to achieve convergence of the growth rate  is very large (of the order of 1000 species). It is clear that 
the discrete power-law method would be highly beneficial in this case.

The only other method incorporating dust dynamics and grain size evolution in hydrodynamical simulations is by 
\citet{mckinnon18} who use a piecewise linear method in the moving-mesh code AREPO.  They demonstrate its applicability on determining the contribution of different grain processes, such as sputtering, shattering, coagulation and dust growth on setting the dust grain size distribution. However, they do not investigate the resolution of the dust distribution on the results.  In Paper~I  we showed that the piecewise power-law method needs significantly less bins than the piecewise linear method to achieve the same accuracy. With a minimum of 4 discrete bins for just modelling grain sputtering using the power-law method, this means 
that it is unclear whether the piece-wise linear method is useful in practise. In a subsequent paper we will investigate
the additional effect of grain shattering and vaporisation on the dust-distribution throughout a C-type shock and 
compare it to the piecewise linear method.

\section*{Acknowledgements}
We thank the referees for their comments that improved the manuscript, Tom Hartquist for the useful discussions on grain physics and Paola Caselli for her continued interest and support. RS thanks the  STFC  and the Center for Astrochemical Studies at the Max Planck Institute for Extraterrestrial Physics for funding this PhD project. SVL is supported by a STFC consolidated grant. 
The data presented in this paper is available in the Research Data Leeds Repository, at   https://doi.org/10.5518/1095.



\bibliographystyle{model2-names}
\bibliography{refs} 




\appendix

\section{Grain charge calculation}\label{app:A}
The calculation of the equilibrium grain charge is modified with regard to the previous models of \citet[][2013]{VanLooetal2009}, where grains are always assumed to be negatively charged.  In this paper we relax that assumption.

Our strategy of calculating is similar to the one of \citet{DraineSutin1987} and \citet{Havnesetal1987}. For average grain charges 
$\bar{Z}_g$ of order unity, we calculate the equilibrium charge probability distribution $f(Z_g)$ using 
\begin{equation}
\Gamma_e(Z_g) f(Z_g) + \Gamma_i(Z_g) f(Z_g-1) = 0,
\end{equation}
where $\Gamma_e(Z_g)$ is the current of electrons and $\Gamma(Z_g)$ the current of ions onto grains carrying a charge of $Z_g$
and, then, calculating the average from 
\begin{equation}
\bar{Z}_g = \sum_{Z_g}{Z_g f(Z_g)}.
\end{equation}
On the other hand, for larger values of $|\bar{Z}_g|$, we use 
\begin{equation}\label{eq:averageZg}
\Gamma_e(\bar{Z}_g) + \Gamma_i(\bar{Z}_g) = 0. 
 \end{equation}
 For both calculations we use the currents as expressed by \citet{Havnesetal1987}.
 If $Z_g \leq 0$, the expressions are
 \begin{eqnarray}\nonumber
   &\Gamma_{i}(Z_g) = \pi a^2 n_i  e \frac{c_i^2}{2v_{ig}} 
     \Bigl\{\left(1+2\left(\frac{v_{ig}}{c_j}\right)^2 - 2Z_g\xi_i \right) \\
    &\times {\rm erf} \left(\frac{v_{ig}}{c_i}\right) + \frac{2(v_{ig}/c_i)}{\sqrt{\pi}} 
	\exp\left(-\left(\frac{v_{ig}}{c_i}\right)^2 \right) \Bigr\},
\end{eqnarray}
and
\begin{equation}
   \Gamma_{e}(Z_g) = -\pi a^2 n_e  e \sqrt{\frac{8k_bT_e}{\pi m_e}} \exp(Z_g \xi_e),
\end{equation}
while, for $Z_g > 0$, we use 
\begin{equation}
   \Gamma_{i}(Z_g) = \pi a^2 n_i  e \sqrt{\frac{8k_bT_i}{\pi m_i}} \exp(-Z_g \xi_i),,
\end{equation}
and
\begin{equation}
   \Gamma_{e}(Z_g) = -\pi a^2 n_e  e \sqrt{\frac{8k_bT_e}{\pi m_e}} (1 + Z_g \xi_e),
\end{equation}
where $a$ is the grain radius, $n_i$ the ion number density, $n_e$ the electron number density, $v_{ig}$ the relative 
drift velocity between the grain and the ions, $c_i$ the sound speed of the ions and 
$\xi = e^2/ak_BT$. Note that the expression for $\Gamma_i(Z_g > 0)$ implies that the relative velocities 
between the grains and the ions are smaller than the ion sound speed. This assumption is made 
because it significantly simplifies the calculation of the grain charge.

Using the expression of the current, Eq.~\ref{eq:averageZg} can be expressed as 
\begin{equation}\label{eq:transcendental}
\alpha + \beta \phi  - \exp(-\phi) = 0,
\end{equation}
where $\phi = |Z_g| \xi$ for both negative and positive values of $Z_g$ and $\alpha$ and $\beta$ are positive constant that depend on 
$n_i$, $n_e$, $T_i$, $T_e$ and $v_{gi}$. This  transcendental equation is similar to the one obtained by \citet{Spitzer1941}
and can be easily solved using Halley's method which has cubic convergence (while Newton's method is only quadratic). 
Then, for each different grain size we can calculate its grain charge. However, as a plasma is charge-neutral, that is  $-n_e + n_i + \sum_{j}{Z_{g,j}n_j} = 0$ and the dust grains in the ISM are predominantly negatively charged,  this can only be achieved 
if the electron density is a free parameter. Consequently, the equilibrium condition is only found by iteratively calculating
the grain charges and then the electron density $n_e$, until all converge. Note that Eq.~\ref{eq:transcendental} only has a root 
if $\alpha < 1$ which in effect puts a lower limit on $n_e$. If we ignore the relative velocity of the ions 
and grains or $v_{gi} =0$, we find that this corresponds to 
\begin{equation}
n_{e,min} \approx  n_i \sqrt{\frac{T_i}{T_e} \frac{m_e}{m_i}}.
\end{equation}
When this situation occurs, the small grains in the distribution are negatively charged, while the larger grains become neutral or
positively charged, as can be seen, for example, in Fig.~\ref{fig:adv_6_all}.

\section{Charged particle velocities}\label{app:B}
In a weakly-ionised plasma the momentum equations for a charged species $i$ reduce to 
\begin{equation}\label{eq:reduced}
\alpha_i \rho_i ({\bf E} + {\bf v}_i \times {\bf B}) + \rho_i \rho_n K_{in} ({\bf v}_n - {\bf v}_i) = 0,
\end{equation}
where $\rho_i$ is the density of the charged particles, $\rho_n$ the neutral density, $\alpha_i$ the 
charge-to-mass ratio of the charged species, $K_{in}$ the collision coefficient between the 
neutrals and the charged species, ${\bf v}_i$ the velocity of the  charged species, ${\bf v}_n$
the neutral velocity, $\bf{E}$ the electric field and $\bf{B}$ the magnetic field.
Combined with Ohm's law,
\begin{equation}\label{eq:ohm}
{\bf E} = -{\bf v}_n \times {\bf B} + r_0 \frac{{\bf J} \cdot {\bf B}}{B^2} {\bf B} + r_1 \frac{{\bf J} \times {\bf B}}{B} 
- r_2 \frac{({\bf J} \times {\bf B})\times {\bf B}}{B^2},
\end{equation}
with $r_0$ the resistivity along the magnetic field, $r_1$ the Hall resistivity, $r_2$ the ambipolar resistivity and
${\bf J}$ the current, we can derive the charged velocities. 

It is often easier to work in the frame co-moving with the neutral fluid. Using ${\bf u}_i = {\bf v}_i - {\bf v}_n$ we can rewrite 
Eq.~\ref{eq:reduced} as
\begin{equation}
({\bf E}^* + {\bf u}_i \times {\bf B}) - \frac{B}{\beta} {\bf u}_i = 0,
\end{equation}
where we also substituted in the Hall parameter of the charged species $\beta_i = \alpha_i B/\rho_n K_{in}$ and ${\bf E}^* = {\bf E} + {\bf v}_n \times {\bf B}$ is 
the electric field in the neutral frame. This equation can also be expressed in matrix form as
\begin{equation}\label{eq:matrix}
\left(\begin{array}{ccc} \displaystyle \frac{B}{\beta_i} &  \displaystyle - B_z & \displaystyle B_y\\ \displaystyle B_z & \displaystyle  \frac{B}{\beta_i} & 
\displaystyle -B_x \\ \displaystyle -B_y & \displaystyle B_x  &\displaystyle  \frac{B}{\beta_i}  
\end{array}\right)\left(\begin{array}{c} u_{x,i} \\ u_{y,i} \\ u_{z,i} \end{array}\right) =  \left(\begin{array}{c} E^*_x \\ E^*_y \\ E^*_z \end{array}\right).
\end{equation}
It is then straightforward to find the charged velocities from 
\begin{equation}
\left(\begin{array}{c} u_{x,i} \\ u_{y,i} \\ u_{z,i} \end{array}\right) = 
\left(\begin{array}{ccc} \displaystyle \frac{B}{\beta_i} &  \displaystyle - B_z & \displaystyle B_y\\ \displaystyle B_z & \displaystyle  \frac{B}{\beta_i} & 
\displaystyle -B_x \\ \displaystyle -B_y & \displaystyle B_x  &\displaystyle  \frac{B}{\beta_i}  
\end{array}\right)^{-1}  \left(\begin{array}{c} E^*_x \\ E^*_y \\ E^*_z \end{array}\right),
\end{equation}
as long as the matrix is invertible. As the determinant of the matrix equals $\frac{B^3}{\beta_i} \left(1 + \frac{1}{\beta_i^2}\right)$, we find that this
is satisfied as long as $\beta_i$ is not too large. However, this cannot be guaranteed in 
simulations making  the numerical scheme unstable.

In order to avoid this from occurring, a different choice of coordinate frame is necessary. We set the $z$-axis along the magnetic field and 
the $x$-axis along the current perpendicular to the magnetic field, i.e. ${\bf B} = (0, 0, B)$ and ${\bf J} = (J_{\perp}, 0, J_{||})$. This simplifies 
the expression for the electric field to ${\bf E}^* = \left(r_2 J_{\perp}, -r_1 J_{\perp}, r_0J_{||}\right)$ and Equation~\ref{eq:matrix} to 
\begin{equation}
\left(\begin{array}{ccc} \displaystyle \frac{B}{\beta_i} &  \displaystyle - B & \displaystyle 0\\ \displaystyle B & \displaystyle  \frac{B}{\beta_i} & 
\displaystyle 0 \\ \displaystyle 0 & \displaystyle 0  &\displaystyle  \frac{B}{\beta_i}  
\end{array}\right)\left(\begin{array}{c} u_{x,i }\\ u_{y,i} \\ u_{z,i} \end{array}\right) =  \left(\begin{array}{c} E^*_x \\ E^*_y \\ E^*_z \end{array}\right).
\end{equation}
We can see that the velocity along the magnetic field can be calculated easily as 
\begin{equation}\label{eq:uz}
u_{z,i} = \frac{\beta_i}{B} E^*_z = r_0 \frac{\beta_i}{B} J_{||},
\end{equation}
while the velocities perpendicular to the magnetic field can be found from
 \begin{equation}
\left(\begin{array}{cc} \displaystyle \frac{B}{\beta_i} &  \displaystyle - B \\ \displaystyle B & \displaystyle  \frac{B}{\beta_i}
\end{array}\right)\left(\begin{array}{c} u_{x,i} \\ u_{y,i}  \end{array}\right) =  \left(\begin{array}{c} E^*_x \\ E^*_y \end{array}\right).
\end{equation}
As the determinant of the matrix equals $\displaystyle B^2\left(1 + \frac{1}{\beta_i^2}\right)$, this matrix is unconditionally invertible
and we find 
\begin{equation}\label{eq:ux}
u_x = \frac{\beta_i^2}{1 + \beta_i^2} \left(\frac{r_2}{\beta} - r_1\right) \frac{J_{\perp}}{B},
\end{equation}
and
\begin{equation}\label{eq:uy}
u_y = -\frac{\beta_i^2}{1 + \beta_i^2} \left( r_2  + \frac{r_1}{\beta_i} \right) \frac{J_{\perp}}{B}.
\end{equation}
This solution shows that, for $\beta_i \rightarrow \infty$, $u_{x,i} = r_1J_{\perp}/B$ and $u_{y,i }= r_2 J_{\perp}/B$, while, for $\beta_i \rightarrow 0$,
both $u_{x,i}$ and $u_{y,i}$ are zero and the charged particles move with the neutral, as expected. Furthermore, using ${\bf J} = \sum_i{\alpha_i \rho_i {\bf u}_i}$ and substituting the charged velocities from Eqs.~\ref{eq:uz}-\ref{eq:uy}, we find that ${\bf J} = (J_{\perp}, 0, J_{||})$, as required.
Thus, this way of deriving the charged velocities is robust although it does requires extra calculations to rotate the frame of reference, first, along the local magnetic field  and then back to the original frame.


\label{lastpage}
\end{document}